# Identifying Growth-Patterns in Children by Applying Cluster analysis to Electronic Medical Records

Moumita Bhattacharya[1], Deborah Ehrenthal, MD, MPH[4,5], Hagit Shatkay, PhD[1,2,3,6]
[1]Computational Biomedicine Lab, Computer and Information Sciences, University of Delaware, Newark, DE, USA
[2]Center for Bioinformatics and Computational Biology,
Delaware Biotechnology Institute, University of Delaware, Newark, DE, USA
[3]Dept. of Biomedical Engineering, University of Delaware, Newark, DE, USA
[4]Department of Obstetrics and Gynecology, Population Health Sciences and Medicine,
School of Medicine and Public Health, University of Wisconsin-Madison, WI, USA
[5]Department of Obstetrics and Gynecology, Christiana Care Health System, Wilmington, DE, USA
[6]School of Computing, Queen's University, Kingston, ON, K7L 3N6, Canada
moumitab@udel.edu, ehrenthal@wisc.edu, shatkay@udel.edu

*Abstract*— **Obesity is one of the leading health concerns in the United States. Researchers and health care providers are interested in understanding factors affecting obesity and detecting the likelihood of obesity as early as possible. In this paper, we set out to recognize children who have higher risk of obesity by identifying distinct growth patterns in them. This is done by using clustering methods, which group together children who share similar body measurements over a period of time. The measurements characterizing children within the same cluster are plotted as a function of age. We refer to these plots as *growth-pattern curves*. We show that distinct growth-pattern curves are associated with different clusters and thus can be used to separate children into the topmost (heaviest), middle, or bottom-most cluster based on early growth measurements.**

## I. INTRODUCTION

According to the Center for Disease Control and Prevention (CDC) and the National Institutes of Health (NIH), the number of obese children has more than doubled over the past 30 years, while that of adolescents has quadrupled [1]. Childhood obesity has both immediate and long term effects on health and well-being. In this paper, we set out to recognize children who have higher risk of obesity by identifying distinct growth patterns in early childhood. This is done by using clustering methods, which group together children who share similar measures over a period of time. Measurements of children within a cluster are then plotted as a function of age. We refer to these plots as *growth-pattern curves*. Here, we show that growth-pattern curves associated with children belonging to different clusters give rise to distinct groups of children, who can be separated into the topmost (heaviest), middle, or bottom-most cluster. Notably, the growth-pattern curves display the growth trajectory of individuals or of a group of individuals; as such they allow for clear identification of children who demonstrate a tendency toward higher weight and are likely to be at risk of obesity.

In this work, we use Electronic Medical Records (EMR) gathered from 4,638 children; the dataset is further described in section II. We note that, the records do not contain a-priori information of whether a child is obese, overweight, average-weight or underweight compared to the rest of the population, which suggests that unsupervised learning, specifically clustering [2], is a suitable approach for analyzing this dataset. Clustering methods support the identification of groups of children who share similar growth patterns over time, as reflected in their EMR data. Summary-measurements characterizing each of the clusters can then be plotted as growth-pattern curves.

Previous studies have employed several supervised machine learning techniques such as neural networks [5], logistic regression [5, 6], decision trees (C4.5) [5, 8], Naïve Bayes [5], Bayesian networks [5, 8] and others, in order to predict the risk of obesity among children and adults. While each of these studies considered a variety of measurements of children or adults at a specific age, none of them takes into account the temporal nature of the data. In the context of childhood obesity prediction, extensive work [5] by Zhang et al. compared six different data mining techniques with logistic regression. In their study, the data used for training and testing recorded parameters of 16,523 children from birth until the age of three. Likewise, in the Fels Longitudinal Study [6], logistic regression models were fitted to relate adult overweight and obesity to childhood and adolescent Body Mass Index values for 166 male subjects and 181 female subjects. Similar studies have been conducted aiming to recognize factors causing childhood obesity and to use these factors in the prediction of adult obesity [8].

Clustering techniques have been applied to extract useful patterns from medical data, typically aiming to identify patients who share common attributes and hence belong to the same risk group. One such study used cluster ensemble and validation techniques to identify subtypes of pervasive developmental disorders (PDD) [4]. Similarly, cluster analysis has been used for biomedical image analysis [10], study of health conditions such as diabetes [7], cardiac disease [11], and others. The clustering methods used in the aforementioned studies include: Hierarchical clustering [2, 4, 7], K-means clustering [2-4], Density Based clustering [2], and EM-based clustering [2-4].

To the best of our knowledge, the study we present here (which includes 4,638 electronic medical records of children over 13 different visits, yielding a total of 60,294 records) is the largest conducted so far for obesity prediction using machine-learning methods. Given the nature of our dataset and the task at hand, namely uncovering distinct growth patterns from longitudinal data, the application of unsupervised clustering methods to the data is a good fit. Moreover, no previous work has utilized the temporal aspect of the data to study obesity risk in children. Additionally, all previous work on obesity prediction were limited to concrete prediction tasks, and to significantly smaller datasets.

The rest of the paper is organized as follows: Section II describes the dataset, the clustering algorithms and the growth-pattern curves used for this study. Section III presents our experiments and results, while Section IV discusses the results in depth, highlights the findings and proposes future work.

## II. EXPERIMENTAL SETTING

### A. Dataset

The dataset we use is drawn from the Delaware Mother Baby Cohort (DMBC), a prospective cohort of mothers and their children [9]. It was constructed using data from electronic medical records (EMRs) and includes 4,638 mother-infant dyads as previously described. This study was approved by the Institutional Review Board. Body measurements for each child were collected from birth until the age of five over thirteen visits. Two types of measures are collected in the EMR: values that remain constant over time such as *ethnicity* and *sex* and values that change over time, also known as *temporal attributes*, such as *weight, height* and *body mass index* (BMI is the ratio of weight to squared height). In this study, we focus on the central growth indicators, namely height, weight and body mass index.

Only 18% of the children have complete records with no missing values for any of the visits. The remaining records have values missing for either height, weight or BMI. The EMR dataset has a significantly lower number of recorded values at the second visit compared to all other visits. This visit is often missed by many parents due to the fact that the first visit is a comprehensive one, and the second visit often takes place shortly after, just to ensure that all is well. As such, the most pertinent information about the child's growth at this stage is provided by the first and third visit, hence we do not use the second visit data in our analysis. Once visit 2 is removed, 1251 of the children have their weights recorded for each of the remaining 12 visits, similarly 664 children have their heights and BMIs recorded. (We note that height is often not recorded in children who cannot yet stand. Since BMI is a function of both height and weight, it can only be recorded when both height and weight measures are available).

Each child is associated with three 12-dimensional vectors, one for the weight measurements $(w_1, w_2,...,w_{12})$, one for the height $(h_1, h_2,...,h_{12})$ and one for the BMI $(b_1, b_2,...,b_{12})$, where the measure at the $i^{th}$ position is the one recorded during the $i^{th}$ visit. Throughout this paper, we refer to these vectors as the *records* associated with each child. For each of the measurement types (height, weight, and BMI) we cluster the children records separately.

### B. Methods

We use two clustering algorithms for our analysis: K-means using Euclidean Distance [2] and Expectation Maximization (EM)-based clustering using Gaussian Mixture Model [2–4]. We compare the results of the two algorithms and identify groups of children who are consistently assigned to the same cluster by both algorithms. This provides a way to intrinsically validate the two algorithms against each other by assessing variation in the cluster assignment. We also experimented with K-means using Mahalanobis distance but the clusters obtained were not well-separated and stable, hence we do not include the results of those experiments in this paper.

Figure 1 provides an example of a traditional growth chart obtained using our EMR data. Five different percentile values are shown per visit. The topmost plot is obtained by taking each single visit on the x axis, averaging the weight of children who belong to the $90^{th}$ percentile (or above) during that visit, and fitting a curve that connects the mean-weight calculated at this level across the 12 visits. Similarly, the other plots show the $75^{th}$, $50^{th}$, $25^{th}$ and the $10^{th}$ percentile curves. When using traditional growth charts, children may shift from one percentile curve to another between visits. These shifts make it difficult to associate the children with any single growth trajectory, or to compare their growth trajectories to those of other children [1]. In contrast, in this paper, we plot charts of the average attribute values (*weight*, *height* and *body mass index*), taken over all the children belonging to each cluster, as a function of mean age. We refer to these charts as *Growth-pattern curves* to distinguish them from traditional growth charts (see e.g. Figure 2). As our growth-pattern curves are based on measurements of the same group of children over a period of time, we can identify children with distinct growth trajectories and accordingly recognize children at greater risk of obesity.

## III. RESULTS

The clustering algorithms have been separately applied to the records pertaining to each of the temporal attributes. We experimented with several values for *K*, the number of clusters, where *K* ranges from 2 to 5. The graphs shown in Figure 2 display growth-pattern curves of children belonging to clusters obtained using the EM-based clustering algorithm.

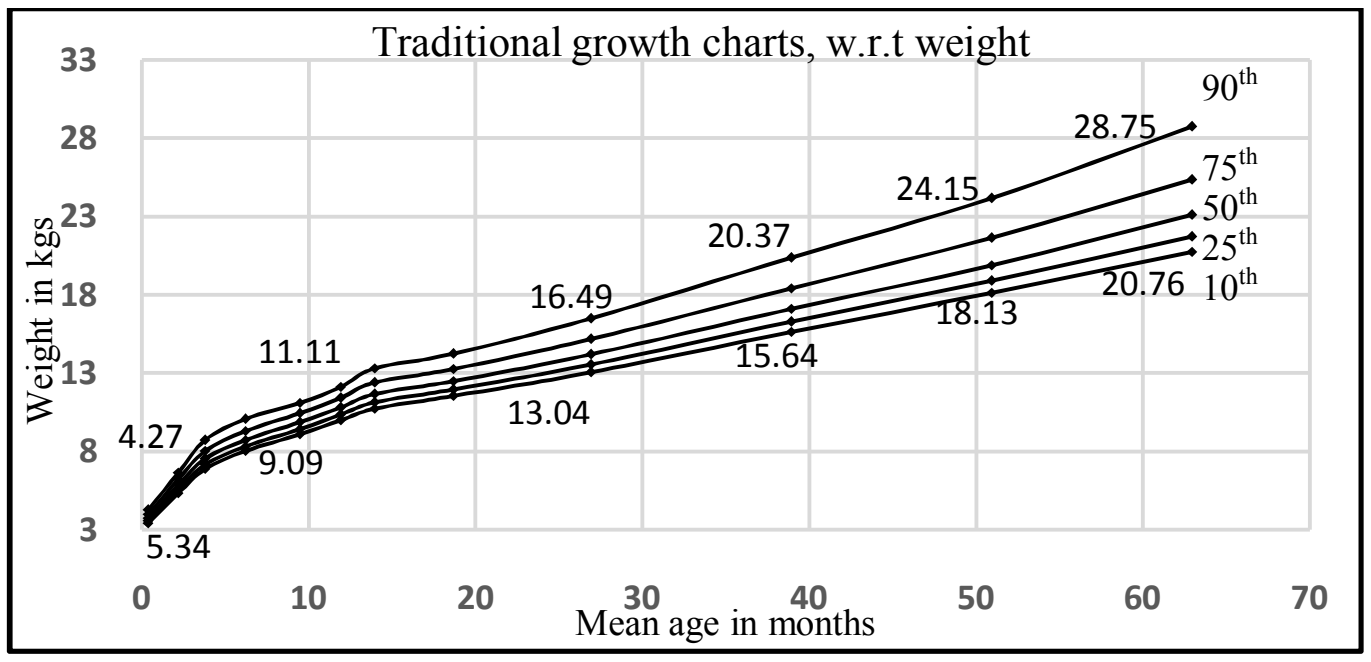

Figure 1. Traditional growth chart obtained from the EMR data with respect to weight. The five charts correspond to the $90^{th}$, $75^{th}$, $50^{th}$, $25^{th}$ and $10^{th}$ percentile curves (from top to bottom respectively).

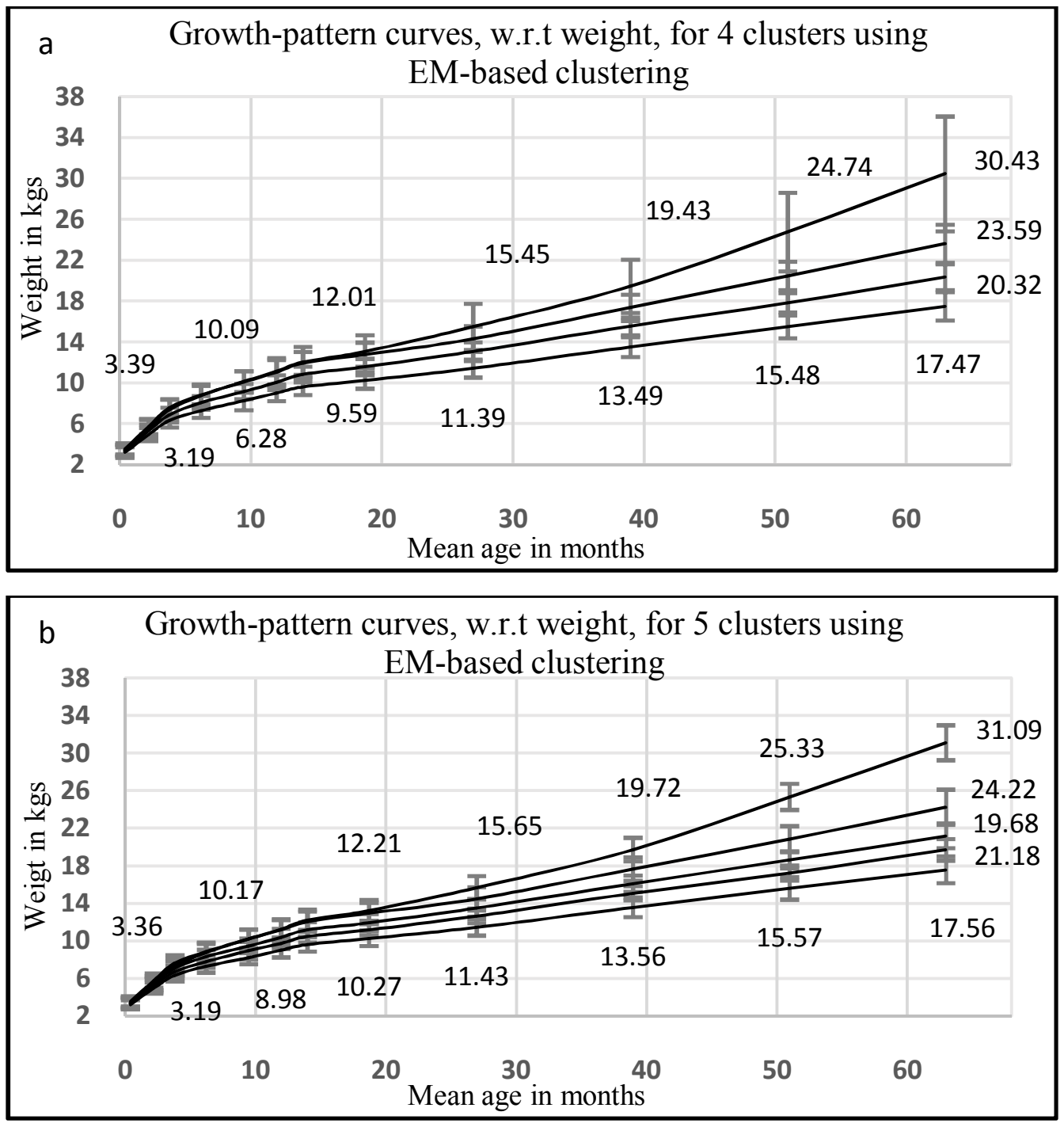

Figure 2. Growth-pattern curves obtained w.r.t weight as a function of mean age for all the children clustered into (a) four and (b) five distinct groups using the EM-based clustering algorithm. The Y-axis shows children weight at each visit in kilograms (kgs) and the X-axis shows mean age in months at each of the twelve visits.

The growth-pattern curves for the different clusters show similar growth rates for all children from birth until the age of 20 months, irrespective of the value of K or the algorithm used to obtain the clusters. In contrast, after the age of 20 months, the growth patterns of children belonging to different clusters diverge when the number of clusters, K, is either 4 or 5.

The growth-pattern curves associated with the clusters obtained from K-means using Euclidean distance also demonstrate similar trends when *K* is either 4 or 5, as can be seen from the graphs shown in Figure 3. When the number of clusters, *K*, is either 4 or 5, we notice that some of the resulting clusters are not well-separated. As such, when *K* is either 4 or 5, we refer to the middle two clusters (i.e. the two clusters below the topmost cluster) together as *the middle cluster*. Moreover, in the 5-cluster scenario we also combine the bottom two clusters and refer to the combination as *the bottom-most* cluster.

We compared the results of the different clustering algorithms to identify groups of children who are consistently assigned to the same cluster by both the algorithms. The motivation for comparing results obtained from different clustering algorithms is to intrinsically validate the consistency of the results, and assess the stability and the tightness of the clusters produced by the two different algorithms. This comparison also gives rise to distinct groups of children, who can be treated separately. Specifically, all the children who are consistently assigned to the topmost cluster can be identified as the group of children who have the highest weight over the period of time, compared to all other children in the dataset.

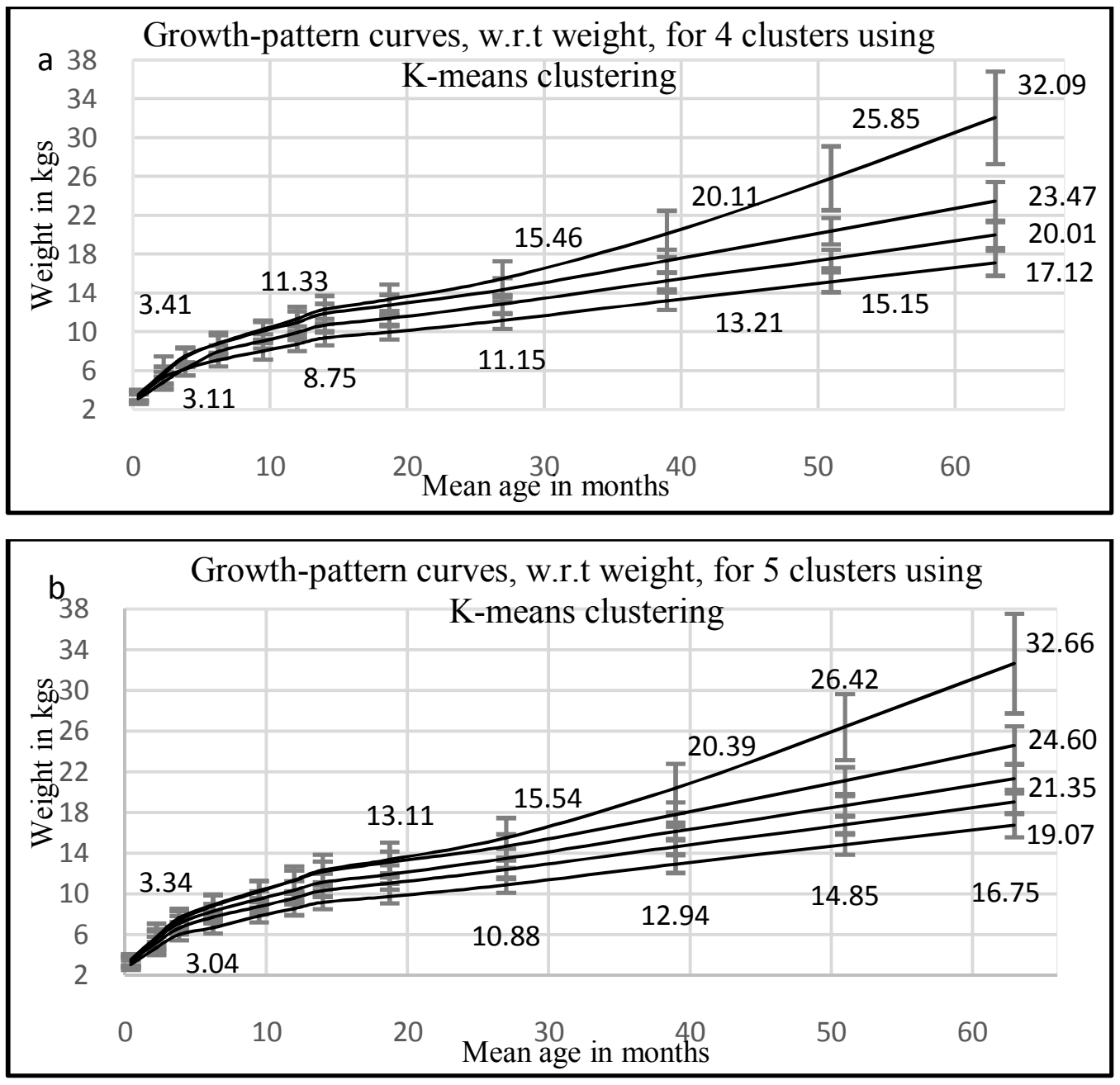

Figure 3. Growth-pattern curves obtained w.r.t weight as a function of mean age for all the children clustered into (a) four and (b) five distinct groups using the K-means clustering algorithm. The Y-axis shows children weight at each visit in kilograms (kgs) and the X-axis shows mean age in months at each of the twelve visits

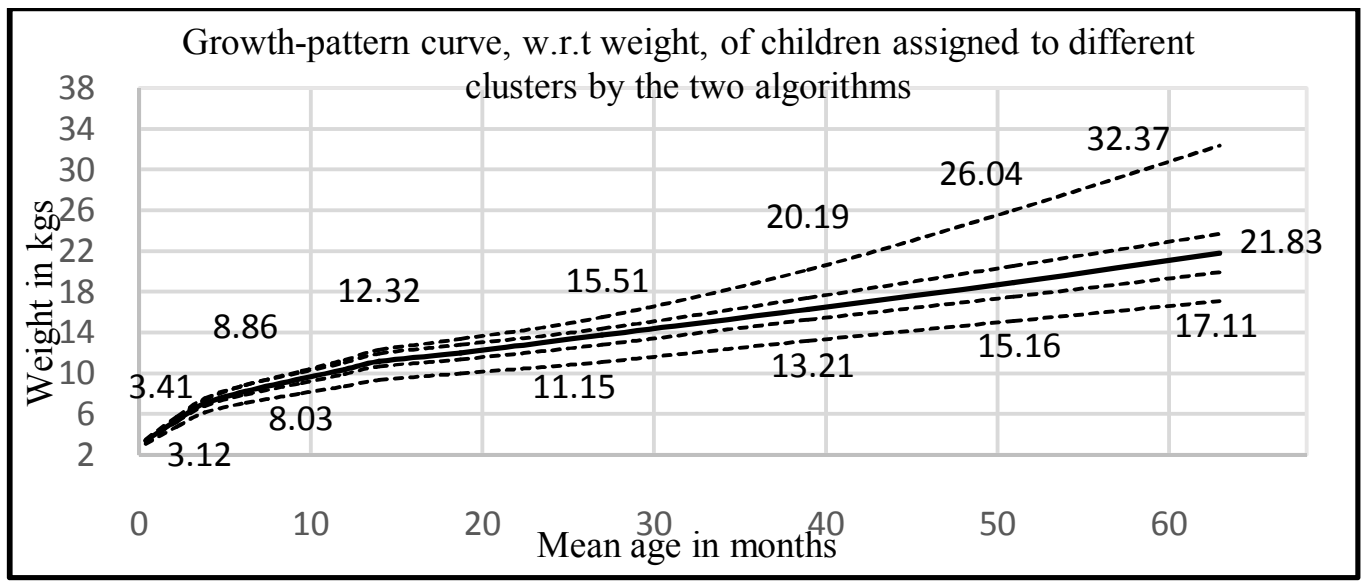

Figure 4. Growth-pattern curve of the 55 children who are assigned to different clusters by the two algorithms (shown as solid line) superimposed on the growth-pattern curves of 1196 children consistently assigned to the topmost, bottom-most and middle two clusters (shown as dashed lines).

Comparing the clusters obtained from the K-means using Euclidean distance with those obtained from the EM-based clustering applied to the weight measurement of 1251 children, we observe that both algorithms assign the same 58 children to the topmost cluster, and both assign 230 children to the cluster below the topmost cluster. Similarly, the bottom-most cluster consists of 356 children, and the cluster above the bottom-most cluster consists of the same 552 children. Of the 1251 children, only 55 children are assigned to different clusters by the two algorithms. Table 1 shows the number of children assigned to each cluster by the two algorithms. The respective growth-pattern curve for the latter 55 children lies between the topmost and the bottom-most clusters, as can be seen from Figure 4. Specifically, the weights of these children are closer to the topmost cluster until the age of 30 months (the 10th visit) and nearer to the bottom-most cluster between the ages of 30 - 70 months (11th and 12th visits). Figure 4 shows the growth-pattern curve for these 55 children superimposed on the growth-pattern curve of the 4 consistent clusters.

TABLE 1. Confusion Matrix showing the number of children assigned to each cluster by the K-means using Euclidean distance and by the EM-based clustering, where C1–C4 denote the four clusters with C1 indicating the bottom-most, C2 and C3 indicating the middle two and C4 incicating the topmost cluster.

| *EM* \ *K-Means* | *C1* | *C2* | *C3* | *C4* |
|---|---|---|---|---|
| ***C1*** | 356 | 13 | 0 | 0 |
| ***C2*** | 4 | 552 | 36 | 0 |
| ***C3*** | 0 | 0 | 230 | 1 |
| ***C4*** | 0 | 1 | 0 | 58 |

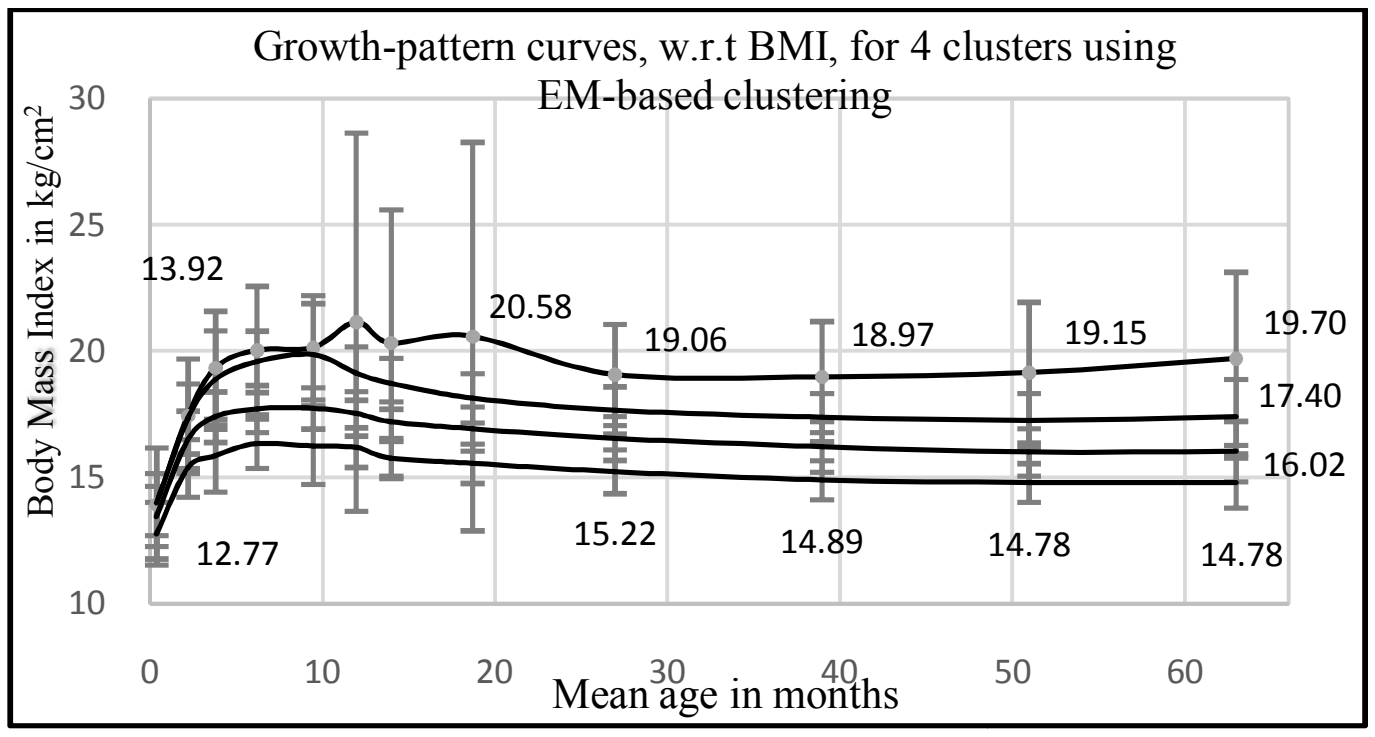

Figure 5. Growth-pattern curves obtained w.r.t BMI as a function of mean age for all the children clustered into four distinct groups using the EM-based clustering algorithm. The Y-axis shows children BMI at each visit in Kilograms per centimeter squared and X-axis shows mean age in months at each of the twelve visits.

The majority of these children are assigned to the topmost cluster by the K-means using the Euclidean distance, but are assigned to lower clusters by the EM clustering algorithm. This can be explained by the observation that the Euclidean distance is a function of the square of the differences along each dimension between the child's record and the cluster centroid. As such, the overall distance calculated significantly increases when the value along some of the dimensions of the record increases. In contrast, the EM-based clustering uses the covariance matrix of each cluster when calculating the probability of a child to belong to a cluster, which takes correlation among different visits into consideration. Thus, the disparity resulting from high values on a few visits is mitigated, and the assignment of the child to a lower-weight cluster remains intact. Hence, the 55 children assigned to higher clusters by the K-means using Euclidean distance are assigned to lower clusters by EM-based clustering.

So far we have discussed clustering with respect to weight using the two algorithms. We conducted similar experiments for clustering with respect to height and body mass index for the 664 children who have those measured at each of the twelve visits. Figure 5 shows the growth-pattern curves obtained with respect to BMI using the EM-based clustering algorithm. The clusters obtained based on BMI are less stable than those obtained with respect to weight. One of the reasons for this instability is that the number of children who had their BMI values during all the twelve visits is almost half the number of children whose weights were measured during all 12 visits. Moreover, the height records are quite noisy, as measurement errors are introduced because children at an early age are typically curled up and need to be stretched in order for their length (height) to be recorded. The noise in the height records also impacts the values of the BMI, since BMI is inversely proportional to the height-squared. Hence, the clustering-signal becomes less clear and discernible.

## IV. CONCLUSION AND FUTURE WORK

Comparing the clusters and their associated growth-pattern curves acquired from the K-means with Euclidean distance and the EM-based clustering algorithms, we obtain distinct groups of children who can roughly be associated with different growth trajectories. Specifically, children belonging to the top-most growth pattern-curve can be tagged as the ones who have the highest weight over a period of time, and hence are likely to be at risk of obesity. Among all the growth-pattern curves associated with different values of *K* obtained through the two algorithms, the ones obtained when the number of clusters is set to 5 using EM-based clustering, yields the tightest and the most distinct topmost cluster. A cursory comparison of our growth-pattern curves with traditional growth charts further demonstrates the utility of our approach and we plan to establish this difference more clearly in future studies. We also plan to examine the impact of the non-temporal attributes on the results of clustering and handle missing data in order to be able to include all the children records in our analysis.

## ACKNOWLEDGEMENT

HS was supported in part by NIH Grant U54GM104941.